\documentclass{ws-procs9x6}

\begin{document}
\newcommand{\mc}{\multicolumn}
\newcommand{\bce}{\begin{center}}
\newcommand{\ece}{\end{center}}
\newcommand{\beq}{\begin{equation}}
\newcommand{\eeq}{\end{equation}}
\newcommand{\bea}{\begin{eqnarray}}

\newcommand{\eea}{\end{eqnarray}}
\newcommand{\cont}{\nonumber\eea\bea}
\newcommand{\cl}[1]{\begin{center} {#1} \end{center}}
\newcommand{\ba}{\begin{array}}
\newcommand{\ea}{\end{array}}
%\newcommand{\arr}{\bea}

% -------------- MATH def -------------------------------
\newcommand{\ab}{{\alpha\beta}}
\newcommand{\cd}{{\gamma\delta}}
\newcommand{\dc}{{\delta\gamma}}
\newcommand{\ac}{{\alpha\gamma}}
\newcommand{\bd}{{\beta\delta}}
\newcommand{\abc}{{\alpha\beta\gamma}}
\newcommand{\eps}{{\epsilon}}
\newcommand{\lam}{{\lambda}}
\newcommand{\mn}{{\mu\nu}}
\newcommand{\mpnp}{{\mu'\nu'}}
\newcommand{\Amuu}{{A_{\mu}}}
\newcommand{\Amuo}{{A^{\mu}}}
\newcommand{\Vmuu}{{V_{\mu}}}
\newcommand{\Vmuo}{{V^{\mu}}}
\newcommand{\Anuu}{{A_{\nu}}}
\newcommand{\Anuo}{{A^{\nu}}}
\newcommand{\Vnuu}{{V_{\nu}}}
\newcommand{\Vnuo}{{V^{\nu}}}
\newcommand{\Fmnu}{{F_{\mu\nu}}}
\newcommand{\Fmno}{{F^{\mu\nu}}}

\newcommand{\abcd}{{\alpha\beta\gamma\delta}}

% Boldmath definitions

\newcommand{\bsigma}{\mbox{\boldmath $\sigma$}}
\newcommand{\btau}{\mbox{\boldmath $\tau$}}
\newcommand{\brho}{\mbox{\boldmath $\rho$}}
\newcommand{\bpipi}{\mbox{\boldmath $\pi\pi$}}
\newcommand{\bss}{\bsigma\!\cdot\!\bsigma}
\newcommand{\btt}{\btau\!\cdot\!\btau}
\newcommand{\bnabla}{\mbox{\boldmath $\nabla$}}
\newcommand{\bphi}{\mbox{\boldmath $\tau$}}
\newcommand{\bvarphi}{\mbox{\boldmath $\rho$}}
\newcommand{\bDelta}{\mbox{\boldmath $\Delta$}}
\newcommand{\bpsi}{\mbox{\boldmath $\psi$}}
\newcommand{\bPsi}{\mbox{\boldmath $\Psi$}}
\newcommand{\bPhi}{\mbox{\boldmath $\Phi$}}
\newcommand{\bnab}{\mbox{\boldmath $\nabla$}}
\newcommand{\bpi}{\mbox{\boldmath $\pi$}}
\newcommand{\btheta}{\mbox{\boldmath $\theta$}}
\newcommand{\bkappa}{\mbox{\boldmath $\kappa$}}

\newcommand{\bA}{{\bf A}}
\newcommand{\bfe}{{\bf e}}
\newcommand{\bb}{{\bf b}}
\newcommand{\br}{{\bf r}}
\newcommand{\bj}{{\bf j}}
\newcommand{\bk}{{\bf k}}
\newcommand{\bl}{{\bf l}}
\newcommand{\bL}{{\bf L}}
\newcommand{\bM}{{\bf M}}
\newcommand{\bp}{{\bf p}}
\newcommand{\bq}{{\bf q}}
\newcommand{\bR}{{\bf R}}
\newcommand{\bs}{{\bf s}}
\newcommand{\bS}{{\bf S}}
\newcommand{\bT}{{\bf T}}
\newcommand{\bv}{{\bf v}}
\newcommand{\bV}{{\bf V}}
\newcommand{\bx}{{\bf x}}
\newcommand{\fph}{${\cal F}$}
\newcommand{\aph}{${\cal A}$}
\newcommand{\dph}{${\cal D}$}
\newcommand{\fpi}{f_\pi}
\newcommand{\mpi}{m_\pi}
\newcommand{\Tr}{{\mbox{\rm Tr}}}
\def\Qb{\overline{Q}}
\newcommand{\delu}{\partial_{\mu}}
\newcommand{\delo}{\partial^{\mu}}
%\newcommand{\half}{{1\over 2}}
%\newcommand{\quart}{{1\over 4}}
%
%
% ------------------ arrow mod ---------------------
\newcommand{\up}{\!\uparrow}
\newcommand{\upup}{\uparrow\uparrow}
\newcommand{\updo}{\uparrow\downarrow}
\newcommand{\uu}{$\uparrow\uparrow$}
\newcommand{\ud}{$\uparrow\downarrow$}
\newcommand{\auu}{$a^{\uparrow\uparrow}$}
\newcommand{\aud}{$a^{\uparrow\downarrow}$}
\newcommand{\pu}{p\!\uparrow}

% ------------------------------------------------------
\newcommand{\qp}{quasiparticle}
\newcommand{\sa}{scattering amplitude}
\newcommand{\ph}{particle-hole}
\newcommand{\qcd}{{\it QCD}}
\newcommand{\integ}{\int\!d}
\newcommand{\ie}{{\sl i.e.~}}
\newcommand{\etal}{{\sl et al.~}}
\newcommand{\etc}{{\sl etc.~}}
\newcommand{\rhs}{{\sl rhs~}}
\newcommand{\lhs}{{\sl lhs~}}
\newcommand{\eg}{{\sl e.g.~}}
\newcommand{\ef}{\epsilon_F}
\newcommand{\sigt}{d^2\sigma/d\Omega dE}
\newcommand{\sige}{{d^2\sigma\over d\Omega dE}}
% ----------------------- ------------------------------
\newcommand{\rpaeq}{\beq
\left ( \begin{array}{cc}
A&B\\
-B^*&-A^*\end{array}\right )
\left ( \begin{array}{c}
X^{(\kappa})\\Y^{(\kappa)}\end{array}\right )=E_\kappa
\left ( \begin{array}{c}
X^{(\kappa})\\Y^{(\kappa)}\end{array}\right )
\eeq}
\newcommand{\ket}[1]{| {#1} \rangle}
\newcommand{\bra}[1]{\langle {#1} |}
\newcommand{\ave}[1]{\langle {#1} \rangle}

%\newcounter{f1}
%\newcounter{f2}
%\renewcommand{\theequation}{\thesubsection.\arabic{equation}}
%\renewcommand{\thetable}{\thesection.\arabic{table}}
\newcommand{\singlespace}{
    \renewcommand{\baselinestretch}{1}\large\normalsize}
\newcommand{\doublespace}{
    \renewcommand{\baselinestretch}{1.6}\large\normalsize}
\newcommand{\bftau}{\mbox{\boldmath $\tau$}}
\newcommand{\bfalpha}{\mbox{\boldmath $\alpha$}}
\newcommand{\bfgamma}{\mbox{\boldmath $\gamma$}}
\newcommand{\bfxi}{\mbox{\boldmath $\xi$}}
\newcommand{\bfbeta}{\mbox{\boldmath $\beta$}}
\newcommand{\bfeta}{\mbox{\boldmath $\eta$}}
\newcommand{\bfpi}{\mbox{\boldmath $\pi$}}
\newcommand{\bfphi}{\mbox{\boldmath $\phi$}}
\newcommand{\bfR}{\mbox{\boldmath ${\cal R}$}}
\newcommand{\bfL}{\mbox{\boldmath ${\cal L}$}}
\newcommand{\bfM}{\mbox{\boldmath ${\cal M}$}}
\def\dblint{\mathop{\rlap{\hbox{$\displaystyle\!\int\!\!\!\!\!\int$}}
    \hbox{$\bigcirc$}}}
\def\ut#1{$\underline{\smash{\vphantom{y}\hbox{#1}}}$}

\def\Pom{{\bf I\!P}}
\def\lsim{\mathrel{\rlap{\lower4pt\hbox{\hskip1pt$\sim$}}
    \raise1pt\hbox{$<$}}}         %less than or approx. symbol
\def\gsim{\mathrel{\rlap{\lower4pt\hbox{\hskip1pt$\sim$}}
    \raise1pt\hbox{$>$}}}         %greater than or approx. symbol
\def\beq{\begin{equation}}
\def\eeq{\end{equation}}
\def\bea{\begin{eqnarray}}
\def\eea{\end{eqnarray}}

\title{Diffractive Hard Dijets and
                     Nuclear Parton Distributions}

\author{I.P. Ivanov$^{a,b)}$, \underline{N.N. Nikolaev}$^{b,c)}$,
W. Sch\"afer$^{d)}$, \\
B.G. Zakharov$^{c)}$, V.R. Zoller$^{e)}$}

\address{$^{A)}$ Novosibirsk State Univeristy, Novosibirsk, Russia\\
$^{B)}$ Institut f. Kernphysik, Forschungszentrum J\"ulich, Germany\\
$^{C)}$ L.D.Landau Institute for Theoretical Physics, Chernogolovka, Russia\\
$^{D)}$ Nordita, Blegdamsvej 17, DK-2100 Copenhagen, Denmark\\
$^{E)}$ Institute for Theoretical and Experimental Physics, Moscow, Russia\\
E-mail: N.Nikolaev$@$fz-juelich.de}

\maketitle

\abstracts{Diffraction plays an exceptional r\^ole in DIS off
heavy nuclei. First, diffraction into hard dijets is an unique
probe of the unintegrated glue in the target. Second, because
diffraction makes 50 per cent of total DIS off a heavy target,
understanding diffraction in a saturation regime is crucial
for a definition of saturated nuclear parton densities. After
brief comments on the Nikolaev-Zakharov (NZ) pomeron-splitting
mechanism for diffractive hard dijet production, I review an
extension of the Nikolaev-Sch\"afer-Schwiete (NSS)  analysis of
diffractive dijet production off nuclei to the definition
of nuclear partons in the saturation regime. I emphasize
the importance of intranuclear distortions of the
parton momentum distributions.}

\section{The Dominance of the Pomeron-Splitting Mechanism for
Diffractive Hard Dijets}

The point that diffraction excitation probes the wave function
of composite systems has been made some 50 years ago by
Landau, Pomeranchuk, Feinberg and Glauber \cite{Landau} - it is
very much relevant to QCD too!

The pQCD diagrams for production of diffractive dijets are shown
in fig~1. In the Landau-Pomeranchuk diagram (b) the limited
transverse momentum $\bp$ of the quark jet comes from the intrinsic momentum
of quarks and/or antiquarks in the beam particle, whereas in
the Pomeron splitting diagram (a) hard jets receive
the transverse momentum from  gluons in the pomeron \cite{NZsplit,NZ92}.
As shown by NSS \cite{NSS} the corresponding diffractive amplitude
is proportional to the unintegrated gluon structure function
of the target proton, ${dG(x,\bp^2)/d\log\bp^2 }$, and the so-called
lightcone distribution amplitude for the beam particle.

\begin{figure}[th]
%\epsfxsize=10cm   %width of figure - will enlarge/reduce the figures
%\epsfbox{JLABdiffr.eps}
%\figurebox{2cm}{2.8cm}{} %to have a box alone
\centerline{\epsfxsize=3in\epsfbox{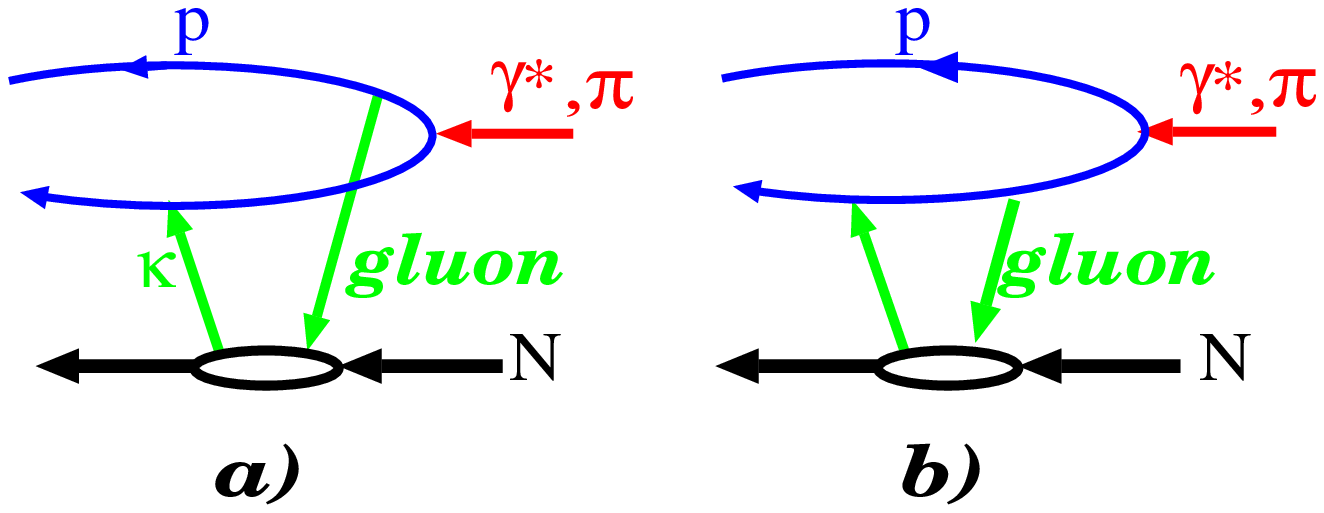}}
\caption{The pQCD diagrams for coherent diffractive dijet excitation}
\end{figure}

The NSS dominance
of the pomeron-splitting contribution for hard dijets
has fully been confirmed by the
NLO order analysis of Chernyak et al. \cite{Chernyak}
and Braun et al. \cite{Regensburg}. The NLO correction to the
NSS amplitude is found to be proportional to the asymptotic
distribution amplitude and  numerically quite substantial, so
that the experimental data by E791 \cite{Ashery} can not
distinguish between the asymptotic and double-humped distribution
amplitudes. According to NSS \cite{NSS} realistic model
distributions do not differ much from the asymptotic one, though. To
my view, the
only caveat in the interpretation of the NLO results is that
the issue of partial reabsorption
of these corrections into the evolution/renormalization of the
pion distribution amplitude has not yet been properly addressed.
Anyway, there emerges a consistent pattern of diffraction of pions
into hard dijets and in view of these findings the claims by Frankfurt
et al. \cite{FMS} that the diffractive amplitude is proportional
to the integrated gluon structure function of the target must be
regarded null and void. Hopefully, some day the E791 collaboration
shall report on the interpretation of their results within the
correct formalism.

The current status of the theory has been comprehensively
reviewed at this Workshop
by Chernyak \cite{JLABChernyak} and Radyushkin
en lieu of Dima Ivanov \cite{JLABIvanov} and there is no point
in repeating the same the third time - the principal
conclusions by NSS have been published some years ago and
are found in \cite{NSS}. I would rather report new results
\cite{INSZZfuture} on the
relevance of diffractive DIS to the hot issue of nuclear
saturation of parton densities.

\section{Diffractive and Truly Inelastic DIS off Free Nucleons and
Heavy Nuclei}

While the above cited NSS papers focused on diffraction
on nuclei in the hard
regime, in the rest of
my talk I would like to discuss the opposite regime
of nuclear saturation. Nuclear saturation is an
opacity of heavy nuclei for color dipole states of
the beam be it a hadron or real, and virtual, photons.
The fundamental point about diffractive
DIS is the counterintuitive result by Nikolaev, Zakharov and Zoller
\cite{NZZdiffr} that for a very heavy nucleus
coherent diffractive DIS in which the target nucleus does
not break and is retained
in the ground state makes precisely 50 per cent
of the total DIS events. Consequently, diffractive DIS is a key
to an understanding of nuclear saturation. I note in passing that
because of the very small fraction of DIS off free nucleons
which is diffractive one, $\eta_D
\lsim
$ 6-10 \%, there is little room for a genuine saturation effects
at HERA. Intuitively, such an importance of diffractive DIS
which can not be treated in terms of parton densities in the
target casts shadow on the interpretation of the saturation regime
in terms of parton densities, which is one of the points from
our analysis \cite{INSZZfuture}.

The
alternative interpretation of nuclear opacity in terms of a fusion
and saturation of nuclear partons goes back to the 1975 papers by Nikolaev
and Valentine Zakharov \cite{NZfusion}: the
Lorentz contraction of relativistic nuclei
entail a spatial overlap of partons with $x \lsim x_{A} \approx
1/R_A m_N$ from different nucleons
and the fusion of overlapping partons results in the saturation
of parton densities per unit area in the impact parameter
space. More recently this idea has been revived in the quantitative
pQCD framework by
McLerran et al. \cite{McLerran}.

\begin{figure}[th]
%\epsfxsize=10cm   %width of figure - will enlarge/reduce the figures
%\epsfbox{JLABdiffr.eps}
%\figurebox{2cm}{3cm}{} %to have a box alone
\centerline{\epsfxsize=3.6in\epsfbox{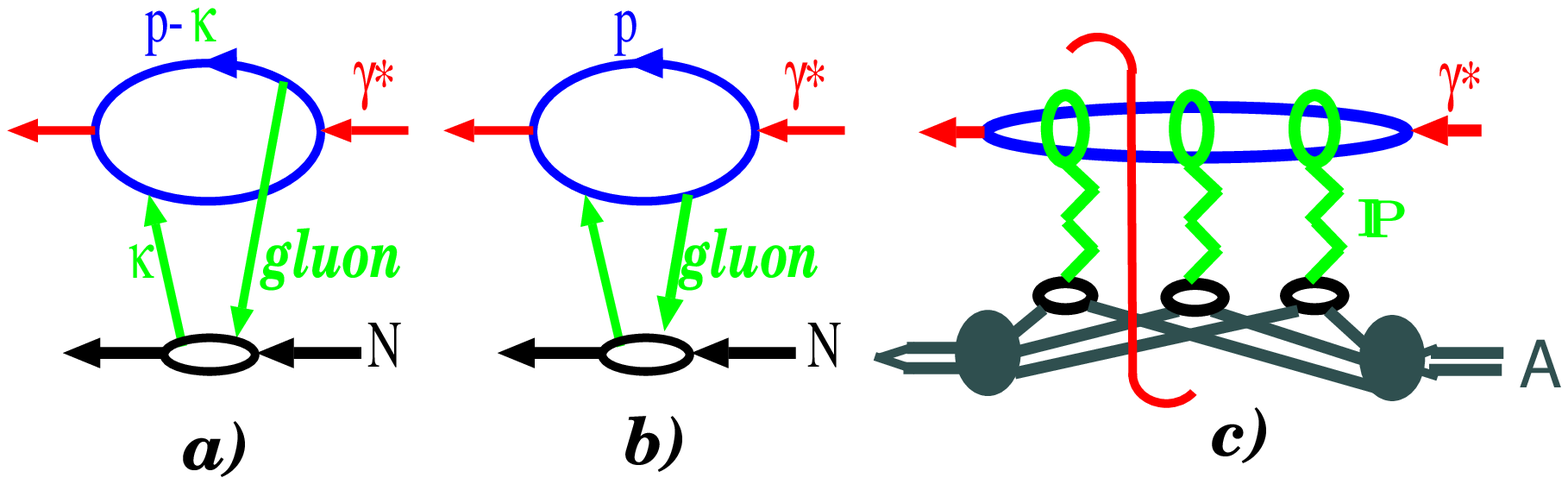}}
\caption{The pQCD diagrams for DIS off protons (a,b) and nuclei (c).
Diagrams (a) and (b) show the 2-gluon tower approximation for
the QCD pomeron. The diagram (c) shows the nuclear multiple scattering
for virtual Compton scattering off nuclei; the diffractive unitarity
cut is indicated. }
\end{figure}

We base our analysis on the color dipole formulation of
DIS \cite{NZ91,NZ92,NZ94,NZZdiffr}.
The total cross section for interaction of the color dipole
$\br$ with the target nucleon equals
\bea
\sigma(r)= \alpha_S(r) \sigma_0\int d^2\bkappa f(\bkappa )\left[1
-\exp(i\bkappa \br )\right]\, ,
\label{eq:2.1}
\eea
where $f(\bkappa )$ is related to the unintegrated glue of the target
by
\bea
f(\bkappa ) = {4\pi \over N_c\sigma_0}\cdot {1\over \kappa^4}
\cdot
{\partial G \over \partial\log\kappa^2}
\label{eq:2.2}
\eea
and is normalized as $\int d^2\bkappa  f(\bkappa )=1. $ Here
$\sigma_{0}$ describes the saturated total cross section for
very large dipoles.  The total virtual photoabsorption cross section
for a free nucleon target equals
\bea
\sigma_N(x,Q^2) = \langle \gamma^*| \sigma(r) |\gamma^* \rangle
=\int_{0}^1 dz \int d^2\br \Psi^*_{\gamma^*}(z,\br) \sigma(\br) \Psi_{\gamma^*}(z,\br)
\\
{d\sigma_N \over d^2\bp dz} =
{\sigma_0\over 2}\cdot { \alpha_S(\bp^2) \over (2\pi)^2}
 \int d^2\bkappa f(\bkappa )
\left|\langle \gamma^*|\bp\rangle - \langle \gamma^*|\bp-\bkappa \rangle\right|^2
\label{eq:2.3}
\eea
where $\bp$ is the transverse momentum, and $z$ the Feynman variable,
of the leading quark in the final state prior the hadronization,
see figs. 2a,b. Notice that the target nucleon is color-excited and
there is no rapidity gap in the final state.
This is a starting point for a definition of the small-x sea generated
from the glue.
The relevant wave functions of the photon are found in \cite{NZsplit,NZ91,NZ92}.

Because of the smallness of the electromagnetic coupling,
the diffractive DIS of fig. 1 amounts to quasielastic scattering of CD
states of the  photon off the target proton \cite{NZ91,NZ92,NZ94}.
In this case the target nucleon is left in the color singlet state
and there is a rapidity gap in the final state.
For the forward diffractive DIS, $\gamma^*p \to (q\bar{q})+p'$, with the
vanishing $(p,p')$ momentum transfer, $\bDelta=0$,
\bea
\left. {d\sigma_D \over d\bDelta^2  dz d^2\bp}\right|_{\bDelta^2=0} =
{1\over 16\pi}\cdot {1 \over (2\pi)^2}
\left|\langle \gamma^*|\sigma(\br)|\bp\rangle\right|^2 \nonumber\\
=
{1\over 16\pi}\cdot {1 \over (2\pi)^2}
[\sigma_0 \alpha_S(\bp^2)]^2
\left|\int d^2\bkappa f(\bkappa)
\left(\langle \gamma^*|\bp\rangle - \langle \gamma^*|\bp-\bkappa \rangle\right)\right|^2
\label{eq:2.4}
\eea
Because $\eta_D$ for a free nucleon target is so small,
in the parton model interpretation of the
proton structure functions one customarily neglects
diffractive absorption corrections , see however warnings in \cite{NZ94}.

Now consider DIS off nuclei at $x\sim x_A$, when interaction of the $q\bar{q}$
states dominates. The coherent
diffractive cross section equals \cite{NZ91,NZZdiffr}
\bea
\sigma_D =\int d^2\bb \langle \gamma^{*} |
\left|1-\exp[-{1\over 2}\sigma(r)T(\bb)]\right|^2
| \gamma^*\rangle \nonumber\\
= \int d^2\bb \int_{0}^1 dz \int {d^2\bp \over (2\pi)^2}
 \left|\langle \gamma^{*} |
\left\{1-\exp[-{1\over 2}\sigma(r)T(\bb)]\right\}|\bp\rangle\right|^2\, .
\label{eq:2.5}
\eea
Here $T(\bb)=\int dz n_{A}(z, \bb)$ is the optical thickness of
a nucleus at an impact parameter $\bb$. The $\sigma_D$ sums
all the unitarity cuts between the exchanged pomerons, so that
none of the nucleons of the nucleus is color-excited and
there is a rapidity gap in the final state, see fig. 2c.

The inelastic DIS describes all events in which one or more
nucleons of the nucleus are color-excited and there is no
rapidity gap in the final state. I omit a somewhat tricky
derivation \cite{INSZZfuture} which is based on the technique
developed in \cite{NNNJETP} and cite only the final result
\bea
{d \sigma_{in}\over d^2\bp dz }   =  {1 \over (2\pi)^2}\int d^2\bb
 \int d^2\br' d^2\br
\exp[i\bp(\br'-\br)]\Psi^*(\br')\Psi(\br)\nonumber\\
\left\{\exp[-{1\over 2}\sigma(\br-\br')T(\bb)]-
\exp[-{1\over 2}[\sigma(\br)+\sigma(\br')]T(\bb)]\right\}\, .
\label{eq:2.6}
\eea
The effect of nuclear distortions on the observed momentum
distribution of quarks is obvious: the dependence of nuclear
attenuation factors on $\br,\br'$ shall affect strongly
a computation of the Fourier transform (\ref{eq:2.6}).

Upon the integration over $\bp$ one recovers the familiar
color dipole Glauber-Gribov formulas \cite{NZ91,NZ92,NZZdiffr}
for the inelastic and total cross sections
\bea
\sigma_{in} =\int d^2\bb \langle \gamma^{*} |
1-\exp[-\sigma(r)T(\bb)
| \gamma^*\rangle \\
\label{eq:2.7}
\sigma_A=\sigma_D+\sigma_{in} =2\int d^2\bb \langle \gamma^{*} |
1-\exp[-{1\over 2}\sigma(r)T(\bb)
| \gamma^*\rangle
\label{eq:2.8}
\eea

%------------  section 3

\section{Nuclear Parton Distributions as Defined by Diffraction}

The next issue is whether nuclear DIS
can be given the conventional parton model interpretation or not.
For the evaluation of the inclusive spectrum of quarks in
inelastic DIS we resort to the NSS representation \cite{NSS}
\bea
\Gamma_A(\bb,\br)= 1-\exp\left[-{1\over 2}\sigma(r)T(\bb)\right]
=
 \int d^2\bkappa  \phi_{WW}(\bkappa )[1-\exp(i
\bkappa \br) ] \, .
\label{eq:3.1}
\eea
There is a  close analogy to the representation (\ref{eq:2.1}),(\ref{eq:2.2})
in terms of $f(\bkappa)$ and
\beq
 \phi_{WW}(\bkappa ) =
\sum_{j=1}^{\infty} \nu_A^j(\bb)
\cdot {1 \over j!}  f^{(j)}(\bkappa )
\exp\left[-\nu_A(\bb)\right]
\label{eq:3.2}
\eeq
can be interpreted as the  unintegrated nuclear Weizs\"acker-Williams (WW)
glue per unit area in the impact parameter plane, normalized as
\beq
\int d^2\bkappa \phi_{WW}(\bkappa )= 1-\exp[-\nu_A(\bb)]\, .
\label{eq:3.3}
\eeq
Here $$\nu_A(\bb)=
{1 \over 2}\alpha_S(r)\sigma_0T(\bb)$$ defines the nuclear opacity
and the $j$-fold convolutions
$$
f^{(j)}(\bkappa )= \int \prod_{i}^j d^2\bkappa _{i}
f(\bkappa _{i})
\delta(\bkappa -\sum_{i}^j \bkappa _i)
$$
describe the contribution to the diffractive amplitudes from
the j split pomerons \cite{NSS}.
 The hard asymptotics
of the WW glue has been analyzed by NSS, here I only
mention that broadening of convolutions compensates completely
the nuclear attenuation effects obvious in the expansion
(\ref{eq:3.2}) and, furthermore, leads to a nuclear antishadowing
for hard dijets \cite{NSS}.

A somewhat involved analysis of properties of convolutions in
the soft region shows that they develop a plateau-like behaviour
with the width of the plateau which expands $\propto j$. Here
I only point out that the gross features of WW glue in the soft region
are well
reproduced by
\bea
\phi_{WW}(\bkappa) \approx  {1\over \pi}  {Q_A^2 \over (\bkappa^2 +Q_{A}^2)^2}\, ,
\label{eq:3.4}
\eea
where the saturation scale
$
Q_A^2 =  \nu_A(\bb)  Q_0^2 \propto A^{1/3}\, .
$
The soft parameters $Q_0^2$ and $\sigma_0 $ are related
to the integrated glue of the proton in
soft region,
$$
Q_{0}^2\sigma_0 \sim {2\pi^2 \over N_c} G_{soft}\,, ~~~G_{soft}\sim 1\,.
$$
Making use of the NSS representation (\ref{eq:3.1}) and
the normalization (\ref{eq:3.3}), after some algebra one finds
for the saturation domain of $\bp^2 \lsim Q^2 \lsim Q_A^2$
\bea
{d \sigma_{in}\over d^2\bb d^2\bp dz}  = {1\over(2\pi)^2}  \int d^2\bkappa
\phi_{WW}(\bkappa)
\left|\langle \gamma^* |\bp+\bkappa\rangle \right|^2 \\
\label{eq:3.5}
{d \sigma_{D}\over d^2\bb  d^2\bp dz} = {1\over(2\pi)^2}
\left| \int d^2\bkappa\phi_{WW}(\bkappa^2)
(\langle \gamma^* |\bp\rangle -
\langle \gamma^* |\bp-\bkappa\rangle) \right|^2 \nonumber\\
\approx {1\over(2\pi)^2}
\left| \int d^2\bkappa\phi_{WW}(\bkappa) \right|^2
\left|\langle \gamma^* |\bp\rangle\right|^2
\approx {1\over(2\pi)^2}
\left|\langle \gamma^* |\bp\rangle\right|^2
\label{eq:3.6}
\eea
The last
result is obvious from (\ref{eq:2.5}) because in this case
all the color dipoles in the virtual photon meet the opacity criterion
$\sigma(r)T(\bb)\gsim 1$, so that the nuclear attenuation terms
can be neglected altogether.

%----------  section 4
\section{The interpretation of the results}

Following the conventional parton model wisdom,
one may try defining the nuclear sea quark density per unit area in the impact
parameter space
\bea
{d\bar{q} \over d^2\bb d^2\bp} = {1\over 2}\cdot{Q^2 \over 4\pi^2 \alpha_{em}}
\cdot{d[\sigma_D+\sigma_{in}] \over d^2\bb d^2\bp} =
{1\over 2}\cdot{Q^2 \over 4\pi^2 \alpha_{em}}
\cdot\int dz  \nonumber\\ \times
\left\{ \left| \int d^2\bkappa\phi_{WW}(\bkappa) \right|^2
\left|\langle \gamma^* |\bp\rangle\right|^2 + \int d^2\bkappa
\phi_{WW}(\bkappa)
\left|\langle \gamma^* |\bp+\bkappa\rangle \right|^2\right\}
\label{eq:4.1}
\eea
It is a nonlinear functional of the NSS-defined WW glue of a nucleus.
The quadratic term comes from diffractive DIS and measures the
momentum distribution of quarks and antiquarks in the $q\bar{q}$
Fock state of the photon. It has no counterpart in DIS off
free nucleons because diffractive DIS off free nucleons is negligible
small even at HERA, $\eta_D \lsim $ 6-10 \%. The linear term comes
from the truly inelastic DIS with color excitation of nucleons
of the target nucleus. As such, it is a counterpart of standard
DIS off free nucleons, but as a function of the photon wave
function and nuclear WW gluon distribution it is completely
different from (\ref{eq:2.3}) for free nucleons. This difference
is entirely due to strong intranuclear distortions of the outgoing quark and
antiquark waves in inelastic DIS off nuclei.

Up to now I specified neither the wave function of the
photon nor the spin of charged partons - they could well have been
scalar or spin-1 ones -, nor the color representation for charged partons.
All our results would hold for any weakly interacting
projectile such that elastic scattering is
negligible small and diffraction excitation amounts to
quasielastic scattering of Fock states of the projectile \cite{NZ91,NZ92}.
Now take the conventional spin-${1\over 2}$
partons and the photon's virtuality $Q^2 \lsim Q_A^2$ such that
the opacity criterion is met for all color dipoles of the photon.
Then upon
the z-integration one finds for $\bp^2 \lsim Q^2$ the
plateau-like distribution from diffractive DIS,
\beq
\left.{d\bar{q} \over d^2\bb d^2\bp}\right|_{D} = {N_c \over 4\pi^4}\, .
\label{eq:4.2}
\eeq

The inclusive spectrum of sea quarks from inelastic DIS also
exhibits a plateau, but very different from (\ref{eq:4.2}):
\beq
\left.{d\bar{q} \over d^2\bb d^2\bp}\right|_{in} =
{1\over 2}\cdot{Q^2 \over 4\pi^2 \alpha_{em}}\phi_{WW}(0)
\int^{Q^2} d^2\bkappa \left|\langle \gamma^* |\bkappa\rangle \right|^2
= {N_c \over 4\pi^4}\cdot {Q^2 \over Q_A^2}\, .
\label{eq:4.3}
\eeq
The plateau for inelastic DIS extends up to $\bp^2
\sim Q^2_A$ and this nuclear broadening of momentum
distributions of outgoing quarks is an obvious indicator of
strong intranuclear distortions. Its  height does explicitly
depend on $Q^2$ and for $Q^2 \ll Q_A^2$ the inelastic plateau
contributes little to the transverse momentum distribution of
soft quarks. Still, the inelastic plateau extends way beyond
$Q^2$ and its integral contribution to the spectrum of
quarks is exactly equal to that from diffractive DIS.
The two-plateau
structure of the nuclear quark momentum distributions has
not been discussed before. For $Q^2 \gsim Q_{A}^2$ the inelastic plateau
coincides with the diffractive one, the both extend up to
$\bp^2 \lsim Q_{A}^2$. Here we agree with Mueller \cite{Mueller}. 

At this point I notice that after the formal mathematical
manipulations with the NSS representation, the total nuclear cross
section (\ref{eq:2.8})     can be cast in the form 
\beq 
\sigma_{A}
= \int d^2\bb\int dz \int {d^2\bp\over (2\pi)^2}  \int
d^2\bkappa\phi_{WW}(\bkappa) \left|(\langle \gamma^* |\bp\rangle
- \langle \gamma^* |\bp-\bkappa\rangle) \right|^2 
\label{eq:4.4}
\eeq
which resembles (\ref{eq:2.3}): the $\bp$ distribution evolves from
the WW nuclear glue in precisely the same manner as as in DIS off
free nucleons, which suggests the reinterpretation of the differential
form of (\ref{eq:4.4}) in terms of the nuclear IS parton density.
Furthermore, in the saturation regime the crossing terms can
be neglected, while the remaining two terms would coincide with $\sigma_D$
and $\sigma_{in}$, respectively, giving some support to an extention 
of the parton model wisdom about the equality of the IS and FS 
parton densities to nuclear targets too. One should be aware of some 
caveats, though. First of all, the equality of IS and FS densities comes at
the expense of a somewhat weird equating the
diffractive FS spectrum to  DIS
spectrum from the spectator quark of
fig. 2b and $\sigma_{in}$ with the contribution from the scattered
quark, the both evaluated in terms of the WW nuclear glue. Second, 
 as pointed out above, (\ref{eq:4.4}) implicitly includes the
 diffractive interactions, which make up 50 \% of the FS quark
yield. Hence the thus defined parton density appears to be highly
nonuniversal, recall that the diffractive final states are typical
of DIS and would be quite irrelevant, e.g. in nuclear collisions.
Furthermore, in sharp contrast to the situation on the proton
target, in (\ref{eq:4.4}) in the saturation regime, the dominant
contribution comes from the region of $\bp^2\lsim \bkappa^2$, just
opposite to the at not too small $x$ dominant strongly ordered
DGLAP contribution from $\bkappa^2 \ll \bp^2$.

One can go one step further and consider interactions with the
opaque nucleus of the $q\bar{q}g$ Fock states of the photon.
Then the above analysis can be extended to $x \ll x_A$  and
the issue of the $x$-dependence of the saturation scale
$Q_A^2$ can be addressed following the discussion in \cite{NZ94}.
I only
mention here that as far as diffraction is concerned, the WW glue
remains a useful concept and the close correspondence between
$\phi_{WW}(\bkappa)$ for the nucleus and $f(\bkappa)$ for the nucleon is
retained. The details of this analysis will be published elsewhere
\cite{INSZZfuture}. For the shortage of space I didn't report here
the phenomenological consequences.

\section{Summary and Conclusions}

The NSS representation for nuclear profile function gives a
convenient and unique definition of the WW gluon structure
function of the nucleus from soft to hard region. The
conclusion by NSS that
diffraction into hard dijets off nucleons and nuclei is
dominated by the pomeron splitting mechanism has been
confirmed by NLO calculations.

Coherent diffractive DIS is shown to dominate the inclusive spectrum
of leading quarks in DIS off nuclei. The observed spectrum
of diffractive leading quarks measures precisely the momentum distribution
of quarks in the $q\bar{q}$ Fock state of the photon, the r\^ole of
the target nucleus is simply to provide an opacity. It exhibits
a saturation property and a universal plateau
but its interpretation as a saturated density
of sea quarks in a nucleus is questionable. The inelastic DIS
also gives the plateau-like spectrum of observed quarks, but
with the height that depends on $Q^2$.   Nuclear broadening of
the inelastic plateau is a clearcut evidence for
an importance of intranuclear distortions of the spectrum of
a struck quark.

This work has been partly supported by the INTAS grants 97-30494
\& 00-00366 and the DFG grant 436RUS17/119/01. I'm grateful to
Anatoly Radyushkin and Paul Stoler for the invitation to this
Workshop.

\end{document}